\documentclass[%
 reprint,
 amsmath,amssymb,
 aps,
]{revtex4-2}

\usepackage{dcolumn}% Align table columns on decimal point
\usepackage{bm}% bold math
\usepackage{graphicx}
\usepackage{caption}
\usepackage{subcaption}
\usepackage{graphicx,hyperref}
\expandafter\let\csname equation*\endcsname\relax
\expandafter\let\csname endequation*\endcsname\relax
\usepackage{amsmath}
\usepackage{amssymb}
\usepackage{multirow, makecell, comment} 
\newcommand{\fla}[1]{\begin{flalign}#1\end{flalign}}
\usepackage{braket}
\usepackage{lineno}
\usepackage{lipsum}
\usepackage{xcolor}
\usepackage{soul}
\usepackage{array}

\begin{document}

\preprint{APS/123-QED}
\title{A graphene platform for nano scale coherent interaction of surface plasmons with resonant atomic ensembles 
%(Propagation of "Fractional" plasmonic modes on 2D graphene sheets and their coupling with resonant two-level atomic ensemble)
}.

\author{Ali.A. Kamli$^1$}
\author{S.A.Moiseev$^2$}

\affiliation{$^{1}$ Department of Physics, Jazan University, Jazan Box 114, Saudi Arabia.}
\affiliation{$^{2}$Kazan Quantum Center, Kazan National Research Technical University n.a. A.N. Tupolev-KAI, 10 K. Marx St., 420111, Kazan, Russia,
\\
 E-mails: 
 alkamli@jazanu.edu.sa,
 s.a.moiseev@kazanqc.org}

\date{\today}

\begin{abstract}

We propose a 2D graphene structure containing atomic ensemble as a platform for implementing nanoscale enhanced coherent interactions of plasmonic fields with resonant atomic systems. 
We determine the graphene surface plasmon modes, and the properties of its electromagnetic fields, and emphasize the role of graphene sheet separation on the interaction with atomic systems for various dipole orientations and positions between the graphene sheets.
We analyze the conditions for implementation of coherent interaction of SP mode with resonant atomic ensembles.   
By solving the Maxwell-Bloch equations that govern the resonant interaction of surface plasmons with atoms, we derive the modified area theorem, which makes it possible to identify the most common nonlinear patterns in the behavior of plasmons under the studied conditions.
%We performed the detail analysis of the coherent interactions between SP fields and atoms concerning dipole moment orientations, different mode polarization, even and odd solutions mode to the structure and so on.   
We obtain analytical and numerical solutions of the area theorem, and find the possibility of stable propagation of isolated SP pulses of graphene surface plasmon modes at "fractional" pulse area values relative to $\pi$.
We show that the coherent dynamics of SP fields can be realized in nanoscale design and 
%The possibility of observing the predicted  graphene plasmonic modes is discussed.
we highlight the possibilities of using this scheme of coherent dynamics for implementing compact multimode nanoscale quantum memory and its integration with other quantum devices on the proposed platform.

\end{abstract}

\maketitle

\section{Introduction}

Plasmonics, the propagation, manipulations and control of surface plasmon polaritons, continues to be pivotal in modern photonics technologies \cite{Barnes03,Zayats05,Maier07,Tame2013}. 
Surface plasmon polaritons (SPs) are the electromagnetic modes of the dressed light-matter coupling that takes place at the interface of metal like material media. These SPs have long enjoyed interesting optical properties that made them the focus of intense research. 
An important property of SPs is their ability to confine light into nanosclae regions of space due to the highly confined nature of the fields near interfaces.
Such properties of SP fields are of great interest for the coherent interaction of light with resonant atomic ensembles \cite{Kamli08,Moiseev10,Tan14,Torma2015,Asgar18, Gomez2019,Gu20,Duan22}. 
The SPs field confinement has always been associated with energy losses in media, and besides tunability this has been a major setback for plasmonics.  \cite{Kamli08,Ozbay06,Gram2010}. 
Recently graphene plasmonics has been suggested as a new platform for plasmonics due to the new promising properties of graphene \cite{Geim2011,Novoselov2011,Novoselov2005,Falkovsky2007,Goncalves2016,Koppens2011,Abajo2014}. 

Graphene plasmonics supports both transverse magnetic(TM) and transverse electric (TE) polarized surface plasmons (GSP) that propagate at the graphene interface with a dielectric material \cite{Koppens2011,Abajo2014,Liu2017,Huang2017}. 
GSP are:  (1) more tunable due to graphene electric doping and gating and conductivity which can be varied with many parameters like Fermi energy, temperature, field mode, graphene layer thickness and electron density. 
The GSP can be well tuned to desired frequency using such varied parameters. 
(2) Their propagation wave vector parallel to graphene interface is much larger than the free space wave number. 
The GSP wavelength is thus much shorter than the free space wave length which can confine light into nanoscale regions with (3) highly confined fields near graphene interface. 
(4) GSPs couple strongly to light and (5) produce huge optical enhancements which means that due to their strong interaction with light, the resultant SP field near interface has larger intensities than the exciting fields,
which have already opened up exciting possibilities for applications in the field of surface enhanced Raman scattering (SERS) \cite{Ru2008}.
%The enhancement of light-matter interaction offers an exciting feature for applications in surface enhanced Raman scattering (SERS) \cite{Ru2008}.  
(6) For the present work, it is especially important that GSPs enjoy low losses below the Fermi energy which means they have longer propagation distance at helium temperature \cite{ni2018fundamental},
%which opens up unique opportunities for the realization of coherent interaction of plasmons with resonant atomic ensembles.
which can be 10 and more $\mu$m, that opens up unique opportunities for the realization of coherent interaction of plasmons with resonant atomic ensembles.
These exciting properties make GSP attractive for exploring the interaction of GSP with quantum emitters at the fundamental level and their applications in quantum technologies 
\cite{Huidobro2012,Nikitin2011,
Huang2017,Thanopulos2022,Ferreira2020,Grigorenko2012,cui2021graphene}. 
In particular the nano-optics nature of these GSP modes with localized fields near interfaces make plasmonics viable candidates to generate strong coupling with atoms and thus providing a potential platform to explore coherent light-atom interactions and possible applications that require strong coupling with atoms. 

The resonant interaction of light pulses with coherent atomic ensembles plays an important role in quantum optics and laser physics \cite{Allen75,Scully1997}, and recently it has become important for quantum technologies, especially in the development of quantum memory devices \cite{Lvovsky-NatPhot-2009,Sangouard-RMP-2011,Moiseev_UFN2025}.
In this work we propose a platform for nanoscale coherent interaction of SP fields with resonant atomic ensemble  localized between two 2D graphene sheets.  
%In this work, we propose a nanoscale platform  2D graphene+atomic ensemble platform based  for robust implementation of coherent interaction between SP fields and atomic ensemble.

%In particular we like to address the resonant nonlinear interaction of GSP light pulses with an  resonant coherent atomic ensemble located between two graphene 2D sheets.
The most general approach to studying the coherent light-atoms interactions is based on the well known McCall-Hahn pulse area theorem, which they derived for free space interaction \cite{MacCall69}.
This pioneering theoretical and experimental work  has induced much works devoted to its application to solving various problems of the interaction of light pulses with resonant atomic ensembles.
Among them, for example, we note its application to the study of Dicke superradiance \cite{Gr01},  the interaction of light pulses with three-level \cite{Eberly02,Shch15} and four-level atoms \cite{Gut16}, 
description of various modes of laser generation \cite{Arkh16,Pakhom2023},
photon echo
\cite{Hahn71,Fr71,Urman,Moiseev20,Wang2022},
optical quantum memory protocols \cite{Moiseev23,Moiseev_FF2025}, 
and more recently on plasmonics meta-materials \cite{Moiseev_PRA2024}.

Here, we are interested to explore the nonlinear interaction of graphene SPs with two level coherent atomic ensemble between 2D graphene sheets.  
We demonstrate that the area theorem developed here makes it possible to understand the general and distinctive nonlinear patterns of graphene SPs interaction with an atomic ensemble. 
We note that the highly inhomogeneous spatial structure of the electromagnetic field of graphene sheets greatly complicates the theoretical description of the nonlinear effects of their interaction with atoms, but using the area theorem allows us to overcome these difficulties in finding some exact analytical solutions.
The analytical solutions demonstrate the possibility of forming stable SP pulses, the properties of which are very different from the case of their formation in free space.
We also highlight the role of the graphene structure interfaces on the nonlinear dynamics of the GSP propagation and enhanced interaction with an atomic ensemble, explore their advantages to study the dynamics of such interactions and explore their applications in quantum memory protocols and devices.
In particular, the developed theory is also applied to the formation of quantum memory, which makes it possible to determine its basic properties and the most optimal conditions for its implementation.

\section{Graphene Plasmonic modes}

 Figure 1 shows the system of graphene based plasmonics of interest.  
 It consists of 2D thin layers of graphene of surface conductivity $\sigma$ located at coordinates z=0 and z=L. 
 A nanoscale plate of thickness L has   dielectric constant $\varepsilon_s$, while the upper and lower semi infinite layers have dielectric constants $\varepsilon_1$ and $\varepsilon_2$ . 
 In conventional plasmonics, surface plasmons exist when one of the two dielectric media has a metallic like behavior with negative real part dielectric constant. 
Graphene, however, supports surface plasmons even for constant dielectric functions, as we shall see shortly. 
The graphene structure in Fig.1 supports both transverse magnetic (TM) and transverse electric (TE) polarized electromagnetic modes. 

\begin{figure}
%\begin{center}\vspace{1cm}
\includegraphics[width=1.0\linewidth]{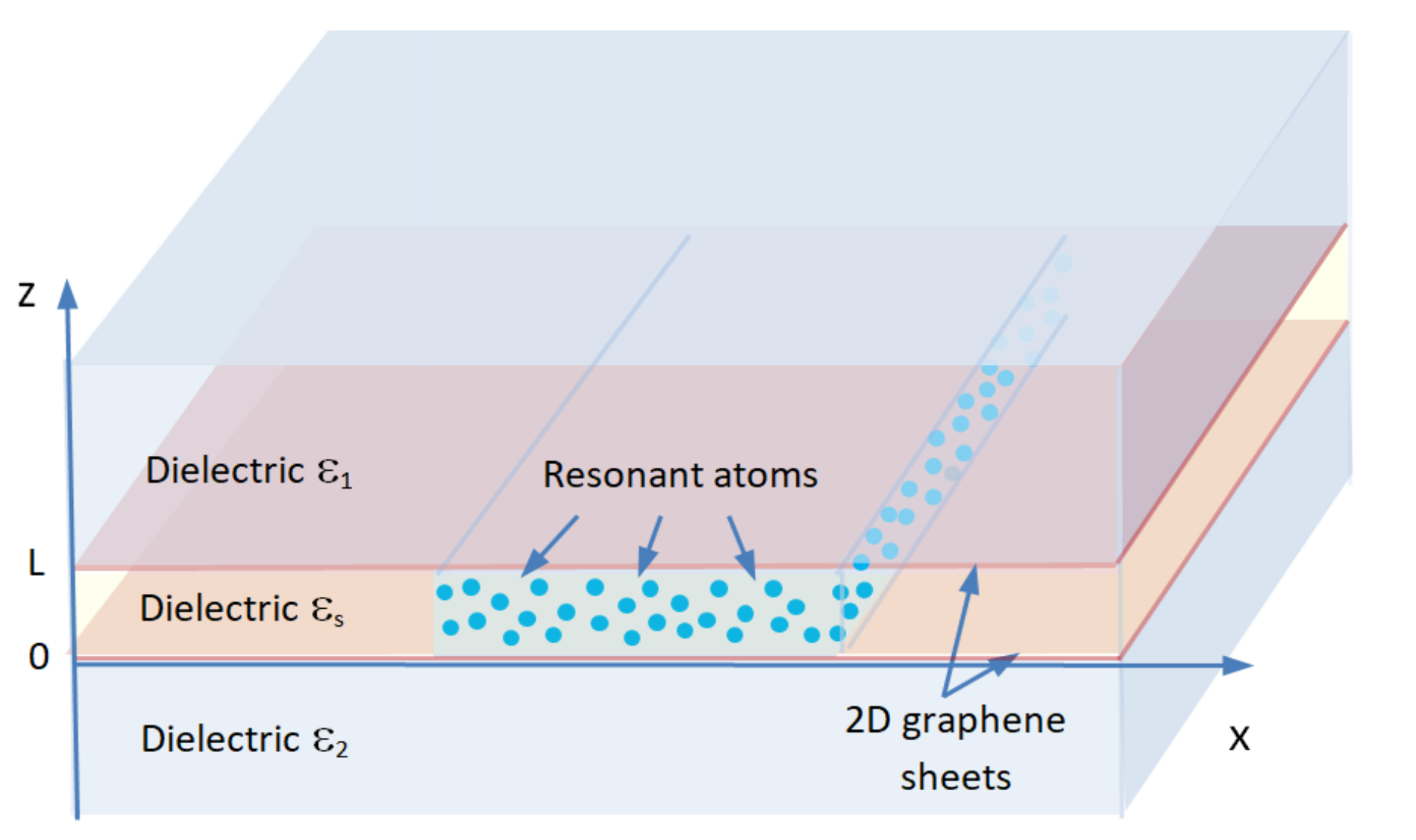}
\captionsetup{justification=justified, singlelinecheck=false}
\caption{ Schematic of 2D two graphene sheets system separated by a dielectric slab of finite nanoscale  thickness $L$ and dielectric constant  $\varepsilon_s$. The upper and lower semi infinite media have dielectric constants $\varepsilon_1$ and $\varepsilon_2$  . 
The two graphene sheets have surface conductivity $\sigma$ and positioned at interfaces $z=0$ and $z=L$. 
The SP pulses propagate along \textbf{x} direction interacting with a two-level inhomogeneously broadened  atomic ensemble located in the middle layer between $z=0, z=L$.}
\end{figure}

These graphene SP modes are confined to interafces at z=0 and z=L,and propagate in the x-direction with complex wave vector $\mathbf{\textit{K}}_{||}$. We shall assume that the propagation along x-direction is facilitated by a channel or groove \cite{Zhang14} 
of width $L_y$ in the y-direction so as to direct the SP propagation along x-direction. 
The SP electric field amplitudes decay away in both sides with distance from the interfaces at $z = 0,L$ with high energy concentration close to interface. 
For TM polarized modes the three layer transverse fields $\mathbf{\textit{E}}$ of frequency $\omega$ satisfy Maxwell wave equations and have the general forms in the outer layers 

\fla{
{\textbf{E}}_(\textbf{r})=& e^{iK_{||}x}
[u(z-L) 
(\hat{\textbf{x}}+\hat{\textbf{z}}\frac{ iK_{||}}{k_1})e^{-k_1(z-L)}A 
\nonumber \\
+& u(-z) 
(\hat{\textbf{x}}-\hat{\textbf{z}}\frac{iK_{||}}{k_2}) e^{k_2 z}D ],
\label{Fields-1}
}

\noindent
and in the middle layer 
\fla{
&{\textbf{E}}(k,z)=e^{iK_{||}x} \cdot
\nonumber
\\
&\left[(\hat{\textbf{x}}+\hat{\textbf{z}}\frac{iK_{||}}{k_s})e^{-k_s z}B+
(\hat{\textbf{x}}-\hat{\textbf{z}}\frac{iK_{||}}{k_2})e^{k_s(z-L)}C\right],
\label{Fields-2}
}

\noindent
the constants $A$, $B$, $C$, and $D$ are to be determined from interface boundary conditions and SP field quantization shortly. 

The wave numbers $K_{||}$ along propagation x-direction, and the normal to interface components $k_j$ are related as  
$k_j\equiv k_j(\omega)=\sqrt{K_{||}^2-(\omega/c)^2\varepsilon_j(\omega)}$
 where the indices (j=1,2 and s) refer respectively to the two outer layers and the middle dielectric slab of thickness L. The wave numbers normal to the interface are characterized by pure imaginary parts so that the SP field amplitudes decay away from interface. The TE modes electric fields can be written in a similar form. Detailed analysis is given here for TM case.

The presence of the graphene two-dimensional charge sheets at the two interfaces presents jump conditions on the tangential component of the magnetic fields in addition to the continuity of the tangential component of the electric field. The condition on continuity of the tangential components of the electric fields at interfaces at z=0, L leads to 
\fla{
&\textbf{E}_1^{||}=\textbf{E}_s^{||},         (z=L),
 \nonumber  \\
&\textbf{E}_2^{||}=\textbf{E}_s^{||},        (z=0),}
\noindent
and the boundary condition on the continuity of the tangential component of the magnetic fields can be written in the form 
\fla{
&\hat{z}\times[\textbf{H}_{1}-\textbf{H}_{s}]=\sigma\textbf{E}_s^{||} ,       (z=L),
\nonumber
\\
&\hat{z}\times[\textbf{H}_{s}-\textbf{H}_{2}]=\sigma\textbf{E}_s^{||} ,        (z=0),
}
\noindent
where $\sigma$  is the graphene two dimensional surface conductivity, given below. Implementing these four boundary conditions leads to the following set of algebraic equations

\fla{
& \textit{A}-\textit{Bf}-\textit{C}=0,
\nonumber
\\
& \textit{B}+\textit{Cf}-\textit{D}=0,
\nonumber
\\
&\textit{A}\left(\frac{\varepsilon_1}{k_1}+\frac{i\sigma}{\varepsilon_0\omega} \right)-\textit{Bf} \frac{\varepsilon_s}{k_s}+ \textit{C} \frac{\varepsilon_s}{k_s}=0,
\nonumber
\\
&\textit{B}\frac{\varepsilon_s}{k_s}-\textit{Cf} \frac{\varepsilon_s}{k_s}+\textit{D} \left(\frac{\varepsilon_2}{k_2}+\frac{i\sigma}{\varepsilon_0\omega} \right)=0, f=e^{-k_{s} L},
\label{Cofficients}
} 
and the solvability condition of \eqref{Cofficients} relates the constants A,B,C and D, and yields the following transcendental equation for the existence of graphene SP modes at the two interface system \cite{Cottam2005,Kamli2014}

\fla{
&\frac{\beta_1+\beta_2}{1+\beta_1\beta_2}=-\text{tanh}(k_{s}L),
\label{Dispersion-1}
}
where 
\fla{
&\beta_{j}=\frac{\varepsilon_{s} k_j}{\varepsilon_{j} k_{s}} \frac{1}{1+(ik_{j}\sigma/\varepsilon_{0}\varepsilon_{j}\omega)}                         (\text{TM modes}),
\nonumber
\\
&\beta_{j}=\frac{k_{j}-i\mu_{0}\omega\sigma}{k_{s}}            (\text{TE modes}),   
}
\noindent
for j= 1,2. 
The condition in \eqref{Dispersion-1} gives the coupled SP modes at any double interface 2D electron gas sheet (2DEGS) \cite{Cottam2005,Kamli2014} that has surface conductivity $\sigma$, for example noble metal and graphene. Here we focus on 2D graphene sheet. The SP coupled modes that exist for small and finite middle layer thickness are designated even and odd for reasons that will become clearer later on when we discuss the electric field profiles. For large sheet separation $(L\rightarrow\infty)$ the two interface modes decouple and we recover the single interface 2DEGS case well known result $\varepsilon_{j}k_{s}+\varepsilon_{s}k_{j}=\sigma k_{j} k_{s}/i\varepsilon_{0} \omega$ for TM case, and $k_{s}+k_{j}=i\sigma\omega/\varepsilon_{0}c^{2}$ for TE case for j=1 or 2. Furthermore by removing the 2D sheet from the interface (conductivity $\sigma=0$), one obtains the single interface SP waves \cite{Cottam2005,Kamli2014}.

The expression for the 2D graphene sheet SP modes given by \eqref{Dispersion-1} is too general for a clear insight into the role of graphene and their interaction with atoms, so we focus on the situation where the two outer dielectric layers are identical i.e $\varepsilon_1=\varepsilon_2=\varepsilon$. In this case we have $k_1=k_2=k$, $\beta_1=\beta_2=\beta$, and the expression in Eq. \eqref{Dispersion-1} splits into two solutions
\fla{
&\beta=-\text{tanh}(\frac{k_{s}L}{2}),
\nonumber
\\
&\beta=-\text{coth}(\frac{k_{s}L}{2}).
\label{Dispersion-2}
}
 The transcendental equations \eqref{Dispersion-1} (or \eqref{Dispersion-2}) are the dispersion relations for the double interface coupled SP modes. 
 They do not in general admit analytic solutions and must be solved numerically to find the wave numbers $K_{||}$  and $k_j$ in terms of graphene SP mode frequency $\omega$. 
 We see from the above equations, that the effect of the 2DEG sheet is quantified by its surface conductivity, which for the graphene sheet of interest here is given by the expressions \cite{Falkovsky2007,Koppens2011,Liu2017}:
\fla{
\sigma=&\sigma_{intra} + \sigma_{inter},
\nonumber
\\
\sigma_{intra} = &i\frac{e^2}{\hbar} 
\frac{2K_{B}T}{\pi\hbar(\omega+i\tau^{-1})} \ln \left(2\text{cosh}[\frac{E_F}{2K_{B}T}]\right),
\nonumber
\\
\sigma_{inter}=&\frac{e^2}{4\hbar}\cdot
[\frac{1}{2}+\frac{1}{\pi}\text{arctan}(\frac{\hbar\omega-E_F}{2K_{B}T})
\nonumber
\\
-&\frac{i}{2\pi}\ln\frac{(\hbar\omega+E_F)^2}{(\hbar\omega-E_F)^2+(2K_{B}T)^2}].
\label{Conductivity}
}

The graphene conductivity in \eqref{Conductivity} receives contributions from the electron transitions within the same band $\sigma_{intra}$, and another part from processes that involve electron transitions across the bands $\sigma_{inter}$. The graphene conductivity is a temperature T dependent, it also depends on the Fermi energy $E_F$, $K_B$   and $\hbar$ are the Boltzmann and Planck constants, $e$ is the electronic charge, $m$ is the electron mass, $\tau$  is SP lifetime ($\tau^{-1}$ is electron scattering rate) and $\omega$ is the excitation frequency.
The graphene conductivity is shown in Fig.2 for both the intraband and interband contributions at room temperature T=300 K. 
For low frequencies the intra band contribution to conductivity is dominant over the inter band part, so in this region the graphene conductivity can be well approximated by the intra band part, while for frequencies near and above Fermi energy the inter band part of conductivity takes over. The negative part in (b) is the onset of inter band contribution and indicates significant losses near and above Fermi energy \cite{Koppens2011,Abajo2014,Liu2017}.  

\begin{figure}   
\centering
     
    \subfloat { {(a)}{\includegraphics[width=8cm]{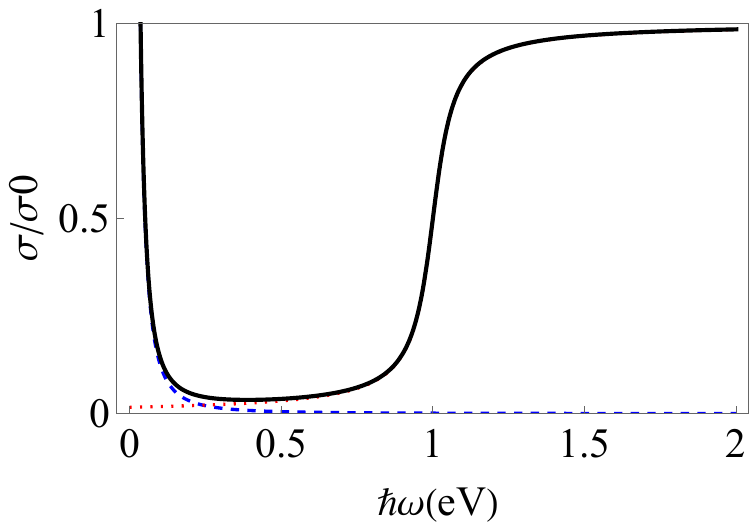} }}
    \qquad
    \subfloat{{(b)} {\includegraphics[width=8cm]{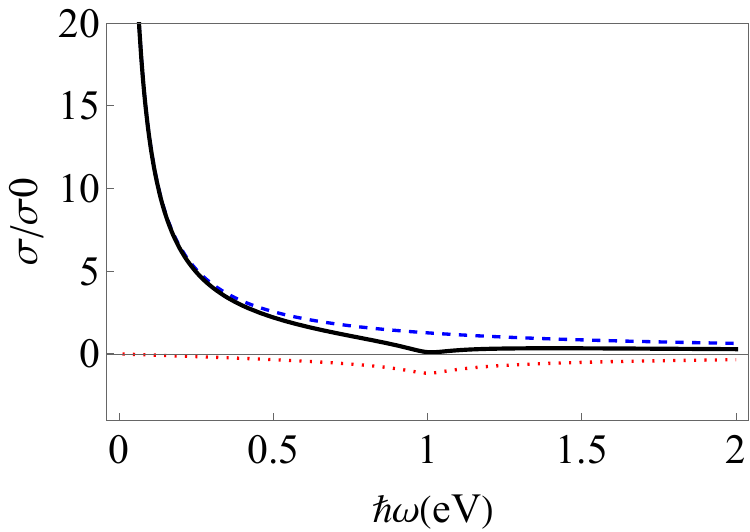} }}
    \caption{Real (a) and imaginary (b) parts of graphene conductivity, where $\sigma_0=e^2/4\hbar$. Blue dashed line is the intra part, red dotted is inter part and solid black is total conductivity. The parameters used are $E_F=1eV$,$T=300K$, $\hbar/\tau=0.001 eV$.}
    \label{fig:example}
\end{figure}

In Fig.3a, we show the wave number behavior as a function of SP mode frequency for the TM tanh solutions as given by condition Eq.\eqref{Dispersion-2} with graphene conductivity as in Eq.\eqref{Conductivity} for the parameters $E_F=1eV$,$K_{B}T=0.02 eV$, $\hbar\tau^{-1}=0.001 eV$, $\varepsilon=1$, and $\varepsilon_s=2$. 
Here we point out a number of observations. 
First of all, we note the wide range of operating mode energy $\hbar\omega=0-0.85 E_F$, available below the Fermi energy, where there is a wide range of frequency tuning, unlike conventional plasmonics.
%First we note there is a wide range of working mode energy $\hbar\omega=0-0.85 E_F$ available below the Fermi energy, so there is a wide range of frequency tunability in contrast to conventional plasmonics. 
In this frequency range, the magnitude of the real part of wave number (black solid line) is  $k_{||}=\text{Re}(K_{||})=10^8-10^9/m$, which is 1-2 orders of magnitude larger than the free space wave number  $(=10^7/m)$ \cite{Koppens2011,Abajo2014,Liu2017}. 
This feature of large wave number is interesting for strong  coupling of SP modes with emitters in the graphene environment. 
Furthermore the imaginary part $\text{Im}(K_{||})$ (blue dotted line) is very small compared to real part, so SP graphene losses can be ignored below Fermi energy at sufficiently small plasmon propagation distances. 
Fig.3b shows the SP propagation length defined as $L_{sp}=1/\text{Im(K)}_{||}$ in units of SP wavelength $\lambda_{sp}=2\pi/\text{Re(K)}_{||}$ as functions of mode energy. 
The propagation length depends on temperature, due to the temperature dependence of graphene conductivity.
%The propagation length is temperature dependent since graphene conductivity is temperature dependent. 
The red dashed and solid blue lines in Fig.3b are the propagation lengths at room temperature T =300 K and at helium temperature T=3 K.
%The dashed red line in Fig.3b is at room temperature T=300K and the solid blue line is at helium temperature T=3K. 
The SP propagation length is two orders of magnitudes larger than the wavelength which translates into propagation length of order $1-10\mu m$. 
It is worth noting that works are currently underway to find methods to increase the propagation length of plasmons in graphene. 
For example, the article \cite{teng2019high} notes the possibility of increasing this length to $15 \mu$m even at room temperature by coating graphene with an additional layer of dielectric.
This feature of low losses will be utilized in next section for SP field quantization and in discussion of quantum memory in later parts of this paper.   

\begin{figure}
    \centering
    \subfloat {(a){\includegraphics[width=8cm]{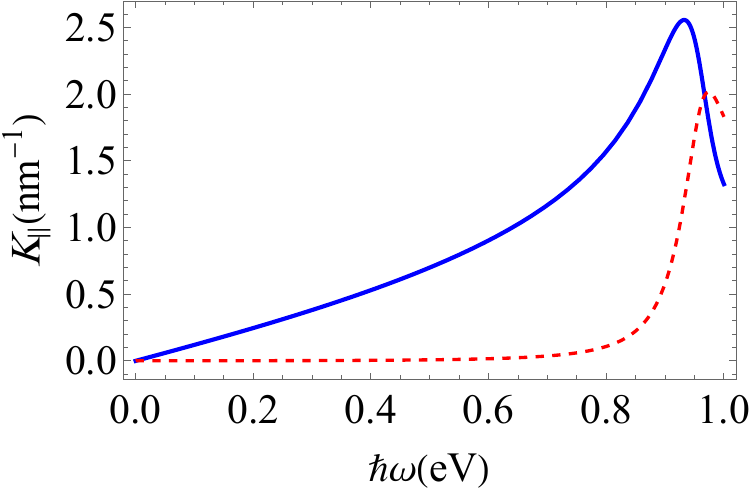} }}
    \qquad
    \subfloat{(b){\includegraphics[width=8cm]{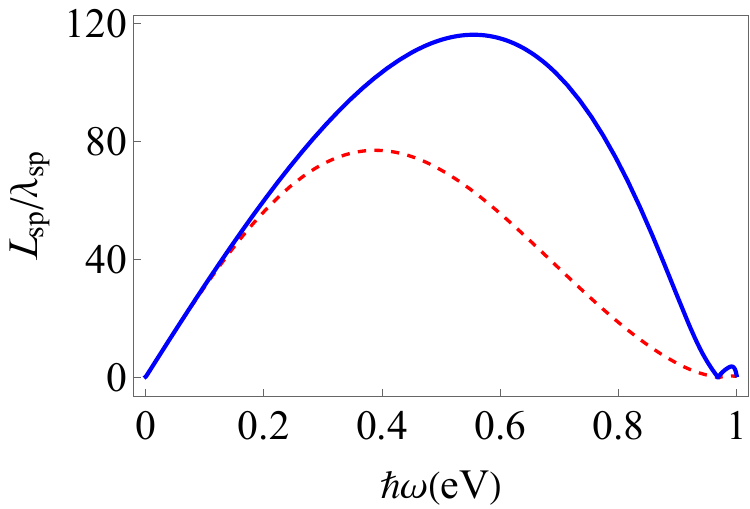} }}
    \caption{(a) Graphene plasmonics complex wave number  as a function of mode energy  for TM mode tanh solutions at room temperature. Solid blue line is $k_{||}=\text{Re}(K_{||} )$, and dotted red line $\text{Im}(K_{||} )$. (b) SP propagation length $L_{sp}=1/\text{Im(K}_{||})$ in units of SP wavelength $\lambda_{sp}=2\pi/\text{Re}(K_{||} )$ as  functions of SP mode energy. Solid blue line for T=3K (helium temp.) and red dashed line for T=300K (room temp.). Parameters used are $E_F=1eV$, $\hbar/\tau=0.001 eV$,$\varepsilon=1$, and $\varepsilon_s=2$. Losses are very low below Fermi energy (1 eV).}
    \label{fig:example}
\end{figure}

\section{Graphene Quantized SP fields }

The graphene SP modes are localized interface eigenmodes that propagate parallel to interface in the x-direction and decay away with increasing distance from interfaces at z=0 and z=L.  We saw in Fig.3 that graphene SP losses are very low for frequency range below the Fermi energy, then in this frequency range losses can be neglected, and the transverse SP fields that satisfy the Maxwell equations are constructed from the Fourier components of SP modes of the type in Eqs. \eqref{Fields-1}-\eqref{Fields-2} as   
\fla{
\hat{\textbf{E}}(\textbf{r},t)=\sum_{\lambda}\int{\textbf{E}_{\lambda}(z,k_{||})e^{ik_{||}(\omega)x-\omega t}dk_{||}}+c.c.,
\label{E-field}
}
\noindent 
%where $L$ is some quantization length and we use $k=k_{||}$ to simplify notations. 

\noindent
where $\lambda(=TM,TE)$ refers to SP modes polarization, and propagation is along x-direction. The localized SP field amplitudes $E_{\lambda}$ in the three layer structure (in Fig.1) are given by the classical mode functions that satisfy wave equations and the boundary conditions as 
\fla{
\textbf{E}_{\lambda}(z,k_{||})=\text{A}_{\lambda}(k_{||})  \textbf{F}_{\lambda}(z,k_{||}),
\label{5}
}
\noindent
where $A_\lambda(k_{||})$ is an overall field amplitude to be determined shortley. The structure functions $F_\lambda(z,k_{||})$ are given for the TM polarized SP modes in the outer layers
\fla{
{\textbf{F}_{TM}}(z,k_{||})=&\text{u(z-L)} 
(\hat{\textbf{x}}+i\hat{\textbf{z}}\frac{ k_{||}}{k})e^{-k(z-L)}+
\nonumber \\
& \text{u(-z)} 
(\hat{\textbf{x}}-i\hat{\textbf{z}}\frac{k_{||}}{k})e^{k z}\zeta,
\label{Exp-electric-field}
}     
where $u(z)$ is the step function, 
and  $\zeta=\frac{\beta-1}{\beta+1}e^{-k_s L}=+1$ or $-1 $, 
with positive sign $(\zeta=+1)$ corresponding to the (symmetric) coth solutions and negative sign $(\zeta=-1)$ to (asymmetric) tanh ones as in \eqref{Dispersion-2}.

In the middle layer $(0<z<L)$ we have 
for the tanh asymmetric solutions $(\zeta=-1)$  

\fla{
&{\textbf{F}_{TM}}(z,k_{||})=2\textit{M} e^{-\frac{1}{2}k_s L}\times
\nonumber \\
&\left[\hat{\textbf{x}}\text{sinhh}[k_s (z-\frac{L}{2})]-\hat{\textbf{z}}\frac{ik_{||}}{k_s}\text{cosh}[k_s(z-\frac{L}{2})]\right],
\label{FTM-1}
}
\noindent
and for the coth symmetric solutions $(\zeta=+1)$.  
\fla{
&{\textbf{F}_{TM}}(z,k_{||}) =2\textit{M} e^{-\frac{1}{2}k_s L} \times
\nonumber
\\
&\left[\hat{\textbf{x}}\text{cosh}[k_s (z-\frac{L}{2})]-\hat{\textbf{z}}\frac{ik_{||}}{k_s}\text{sinh}[k_s(z-\frac{L}{2})]\right],
\label{FTM+1}
}

where
\fla{
{\textit{M}}=\frac{\beta-1}{2\beta}, {\beta}=\frac{\varepsilon_s/k_s}{\varepsilon/k+i\sigma/\varepsilon_0 \omega},
\label{6}
}

The tanh and coth solutions in \eqref{Dispersion-2} correspond to graphene TM polarized SP coupled modes with electric field components parallel to interface are odd (asymmetric) as in \eqref{FTM-1} and even (symmetric) functions of ($\textit{z-L}/2$) as in \eqref{FTM+1}. The tanh and coth solutions are some times referred to as asymmetric(odd) and symmetric (even) solution modes, respectively. 

For TE polarized SP modes, we have in the outer layers  
\fla{
&{\textbf{F}_{TE}}(z,k_{||})=\hat{\textbf{y}}[u(z-L) e^{-k_s(z-L)}-u(-z)\zeta e^{k_sz}], 
\label{field_out}
}
\noindent
and in the middle layer $(0<z<L)$ , 
\fla{
&{\textbf{F}_{TE}}(z,k_{||})=\hat{\textbf{y}} (2\textit{N} e^{-\frac{1}{2}k_s L}) 
\text{sinh}[k_s (z-\frac{L}{2})],   
\label{FTE+1}
}
for $\zeta=+1$, and 
  \fla{
&{\textbf{F}_{TE}}(z,k_{||})=\hat{\textbf{y}} (2\textit{N} e^{-\frac{1}{2}k_s L}) 
\text{cosh}[k_s (z-\frac{L}{2})],  \label{FTE-1}
}

for $\zeta=-1$, where 
\fla{
{\textit{N}}=\frac{1-\nu}{2}, {\nu}=\frac{k_s-i\mu_0 \sigma}{k_s},
\label{N,nu}
}

The above structure functions determine the SP field modes, and we shall now find the field amplitudes from quantization. The quantization procedure is established by recalling that each field mode is an harmonic oscillator of frequency $\omega$, and assigning the field amplitude $A(k_{||})$ the status of quantum harmonic operator. So we make the identification
\fla{
&{\text{A}}_{\lambda}(k_{||})\rightarrow\mathcal{C}_{\lambda}(k_{||}) \hat{a}_{\lambda}(k_{||}) ,{\text{A}_{\lambda}^{*}}(k_{||})\rightarrow\mathcal{C}_{\lambda}^{*}(k_{||}) \hat{a}_{\lambda}^{\dagger}(k_{||}), 
\label{6}
\nonumber
}
where $\hat{a}$ and $\hat{a}^{\dagger}$ are the surface plasmonic bosonic modes annihilation and creation operators that obey the equal time commutation relation
\fla{
&[{\hat{a}_{\lambda}}(k_{||}),{\hat{a}^{\dagger}_{\lambda^{'}}(k_{||}^{'}]=\delta_{\lambda\lambda^{'}} \delta(k_{||}-k_{||}^{'})},
}
and  $\mathcal{C}_{\lambda}(k_{||})$ is a normalization factor to be determined as follows.  We write the SP field energy operator as sum of quantized harmonic modes as
\fla{
&\hat{H}=\frac{1}{2} \int dk_{||}\hbar\omega [\hat{a}_{\lambda}(k_{||})\hat{a}^{\dagger}_{\lambda}(k_{||})+\hat{a}^{\dagger}_{\lambda}(k_{||})\hat{a}_{\lambda}(k_{||})].
\label{Ham-1}
}
On the other hand, the electromagnetic field energy operator in a dispersive but lossless (low losses in the frequency range of interest) dielectric is given by the expression \cite{Novotny2012} as
\fla{
&\hat{H}^{'}=\frac{1}{2}\int d^{3}r \left[\tilde{\varepsilon} |\mathbf{E}(r)|^2+\tilde{\mu}|\mathbf{H}(r)|^2\right],
\label{Ham-2}
 }
 
 \fla{
 \tilde{\varepsilon}=\text{Re}\left[\frac{\partial}{\partial\omega}\left(\omega\varepsilon_0\varepsilon(\omega)\right)\right] , \tilde{\mu}=\text{Re}\left[\frac{\partial}{\partial\omega}\left(\omega\mu_0\mu(\omega)\right)\right],
 \label{dis}
 }

\noindent
where $\varepsilon(z,\omega)$  is the medium dielectric function and $\mu(z,\omega)$  is the magnetic permeability, which are in general position and frequency dependent complex functions, but will be taken constant throughout this work. The magnetic field operator is  calculable from the relation 
%$\textbf{H}=(curl \textbf{E}/i\mu_0\omega)$. 
%$\textbf{H}=\frac{1}{i\mu_0\omega}(\bigtriangledown \times\textbf{E})$. 
$\textbf{H}=\frac{1}{i\mu_0\omega}(\overrightarrow{\bigtriangledown} \times\textbf{E})$. 
Then consistency requires that the two field energy expressions \eqref{Ham-1} and \eqref{Ham-2} are equal. The normalization factor is thus obtained using the SP field operator in Eqs.\eqref{E-field} and its associated magnetic field into Eq.\eqref{Ham-2} and equating the results with the field energy in Eq.\eqref{Ham-1}. This quantization procedure \cite{Tame2013,Kamli08,Ferreira2020} culminates in the following expression for the normalization factor  
\fla{
\mathcal{C}_{\lambda}(\omega)= \sqrt{\frac{\hbar\omega}{ 2\pi \varepsilon_0 L_y L_{\lambda}(\omega)}}.
\label{Norm_factor}
}
For the TM polarized SP fields we have,
\fla{L_{TM}(\omega)=D_{TM}(\omega)+\frac{\omega^2}{c^2}S_{TM}(\omega),
\label{8}
}

\fla{
&\text{D}_{TM}(\omega)= \frac{\text{Re}[\partial_{\omega}(\omega\varepsilon)]}{k}\frac{|k|^2+|k_{||}|^2}{|k|^2} +
\frac{\text{Re}[\partial_{\omega}(\omega\varepsilon_s)]}{k_{s}} e^{-k_{s}L}\times
\nonumber
\\
&(2|M|^2)\left[\frac{|k_{s}|^2+|k_{||}|^2}{|k_{s}|^2}\text{sinh}(k_{s}L) +\zeta \frac{|k_{s}|^2-|k_{||}|^2}{|k_{s}|^2}(k_{s}L)\right],
\label{9}
}

\fla{
&\text{S}_{TM}(\omega)=
\nonumber
\\
&\frac{1}{k} 
|\frac{\varepsilon}{k}|^2+2|M|^2 \frac{e^{-k_s L}}{k_{s}} |\frac{\varepsilon_s}{k_s}|^2  
[\text{sinh}(k_s L)-\zeta (k_s L)].
\label{10}
} 
Similarly for TE polarized SP modes we have 

\fla{L_{TE}(\omega)=D_{TE}(\omega)+\frac{c^2}{\omega^2}S_{TE}(\omega),
\label{8}
}

\fla{
\text{D}_{TE}(\omega)=& \frac{\text{Re}[\partial_{\omega}(\omega\varepsilon)]}{k} +\frac{\text{Re}[\partial_{\omega}(\omega\varepsilon_s)]}{k_s} e^{-k_{s}L}\times
\nonumber
\\
&(2|N|^2)\left[\text{sinh}(k_{s}L) +\zeta k_{s}L\right],
\label{9}
}
 
\fla{
&\text{S}_{TE}(\omega)= \frac{|k|^2+|k_{||}|^2}{k}+2|N|^2\frac{e^{-k_s L}}{k_s}\times,
\nonumber
\\
&\left[(|k_s|^2+|k_{||}|^2) \text{sinh}(k_s L)-(|k_s|^2-|k_{||}|^2)( \zeta k_s L)\right].
\label{10}
}

The normalization factors $\mathcal{C}_{\lambda}$ determine the field amplitudes and are given in terms of various structure parameters. Putting all the pieces together, the quantized electric field operator for the graphene SPs in the two graphene sheet structure can now be rewritten in the more compact form
\fla{
\hat{\textbf{E}}(\textbf{r},t)=&
\sum_{\lambda}\int dk_{||} \mathcal{C}_{\lambda}(k_{||})\textbf{F}_{\lambda}(z,k_{||}) \hat{a}_{\lambda}(k_{||}) e^{i(k_{||} x-\omega t)}
\nonumber
\\
+&H.C.,
\label{Electric-field}
}
 The role of the graphene conductivity enters into quantization through the wave number $k_{||}$ in Eqs.\eqref{Dispersion-1}-\eqref{Dispersion-2} which expllicitly show the dependence of the wave number on graphene conductivity. 
We have considered the quantization process in some details in order to determine the conditions under which the dynamic description of plasmons is valid. 
Namely, below we find the conditions under which these losses can be neglected in the presence of interaction of surface plasmons with an ensemble of resonant atoms.

We emphasize here that the quantization process outlined above is valid in the frequency range where losses are very low and can be ignored.
It is worth noting that this  quantization 
approach has also been used by many authors \cite{Tame2013,Loudon2003,Ferreira2020}. 
Other more elaborate quantization schemes \cite{Huttner92, Matloob1995,Dung1998} are equivalent to this method in situation where losses are very low. More detailed and eloborate quantization of graphene plasmonics will be considered elsewhere.

\section{Pulse propagation and coupling to two level atomic ensemble}

 We are interested to explore the propagation and resonant interaction of short pulses of graphene plasmonics with coherent atomic ensemble inside the nanoscale graphene two-layer structure. 
 The coherent two-level atomic ensemble is inhomogeneously broadened and can be located between two sheets of graphene, as well as above or below the two layer graphene system of Fig.1. 
We also assume that the graphene SP pulse duration is much shorter than the phase relaxation lifetime of the atomic resonant transition ($\delta t_s\ll T_2$),
but sufficiently long  ($\delta t_s \gg \lambda/c$), 
which is a necessary condition for the derivation of the area theorem
\cite{Hughes1998}.
Recent work has shown \cite{Choquette2010} that an atomic ensemble can emit energy in SP modes at least two orders of magnitude stronger than in free space modes. 
Therefore, only the SP modes are included in the Hamiltonian used here in studies of coherent interaction with an atomic ensemble.
The total Hamiltonian of the atomic ensemble and the graphene SP pulses is given as 

\fla{\hat{H}=\hat{H}_a+\hat{H}_f+\hat{H}_{int},
\label{11}
}
with

\fla{\hat{H}_a=&\frac{1}{2}\sum_j\hbar(\omega_0+\Delta_j)\sigma_z^j,
\\
\hat{H}_f=&\frac{1}{2} \int dk_{||}\hbar\omega(k_{||}) [\hat{a}^{\dagger}(k_{||})\hat{a}(k_{||})+H.C],
\\
\hat{H}_{int}=&-\frac{1}{2}\sum_j\int dk_{||}\hbar\mathcal{R}(z_j)\sigma_{+}^j \hat{a}(k_{||}) e^{ik_{||} x_j}+H.C.,
}
where  $\hat{H}_a$  is the atomic ensemble Hamiltonian with central transition frequency $\omega_0$ and detuning $\Delta_j$  for the j-th atom that is inhomogeneously broadened by function $G(\frac{\Delta}{\Delta_{in}})$ with spectral width $\Delta_{in}$,
the two level atom operators are raising $\sigma_{+}^j=\ket{2_j}\bra{1_j}$, lowering $\sigma_{-}^j=\ket{1_j}\bra{2_j}$  operators and the inversion $\sigma_{z}^j=\frac{1}{2}\left(\ket{2_j}\bra{2_j}-\ket{1_j}\bra{1_j}\right)$,
$\hat{H}_f$  is the SP field part with mode frequency $\omega(k_{||})$, 
and $\hat{H}_{int}$  is the dipole interaction Hamiltonian of the SP field with atoms in slowly-varying-envelope approximation,  $r_j=(x_j,z_j)$ is position of j-th atom  and $\hbar$ is the reduced Planck constant. 
The atom - SP field coupling strength $\mathcal{R}(z_j)=2\textbf{d}_{21}^{j}\cdot \textbf{E}(z_j,\omega_0)/\hbar$ is a function of atomic location along z-direction normal to interface and atomic central frequency $\omega_0$ and other structure parameters.
The SP field amplitude is given by Eqs \eqref{Electric-field} and $\textbf{d}_{21}^{j}$ is the dipole moment of the atomic transition. 
It is worth noting the experimental work in which the interaction of surface plasmons with an atomic ensemble with an inhomogeneously broadened resonant transition \cite{Gomez2019} was studied, and the more recent work \cite{Torma2015}, which indicates the possibility of a strong interaction of atoms (molecules) with SP modes under conditions of inhomogeneous broadening of the resonance line.

To explore the transport of SP fields in space, it is more appropriate to work in Heisenberg picture to derive the Maxwell-Bloch equations  of motion of the field and atomic operators \cite{Allen75} by taking into the account coupling constants of atoms with surface plasmon modes   \cite{Moiseev_PRA2024}.

\fla{
&\left( \frac{\partial}{\partial t}+\frac{\gamma_w}{2} + v_g \frac{\partial}{\partial x} \right) \hat{a}(x,t) =
\nonumber
\\
&i\sqrt{\pi/2} \sum_{j} \mathcal{R}^{*}(z_j) \hat{\sigma}_{-}^j(t)\delta(x-x_j)+\sqrt{\gamma_w}\hat{b}_{in}(x,t),
\nonumber
\\
&\frac{\partial \hat{\sigma}_{-}^j(t)}{\partial t} = -  (i\Delta_{j} +\gamma) \hat{\sigma}_{-}^j(t)-\frac{i}{2}\mathcal{R}(z_j)\hat{\sigma}_{z}^j(t)\hat{a}(x_j,t),
\nonumber
\\
&\frac{\partial \hat{\sigma}_{z}^j(t)}{\partial t} = i\left[
\mathcal{R}(z_j)\hat{\sigma}_{+}^j(t)\hat{a}(x_j,t)
-\mathcal{R}^*(z_j)\hat{\sigma}_{-}^j(t)\hat{a}^{\dagger}(x_j,t)\right].
\label{Max_Bloch_eqs}
}

To describe the atomic response to the action of a SP pulse, we introduce a negligibly small decay constant of the atomic coherence $\gamma= \frac{1}{T_2}\ll\Delta_{in},\delta t_s^{-1}$,
$\gamma_w$ and $\hat{b}_{in}(x,t)$ are decay constant of SP mode and the corresponding Langevin force with the following property $\langle \hat{b}_{in}(x,t)\rangle=0$ \cite{Scully1997},
$v_g$ is the SP group velocity calculated from the dispersion relation described above.
 
%$S_{z}(\Delta,x,z,t)=S_{z}(-\Delta,x,z,t)$ 
%and $\gamma\ll\Delta_{in},\delta t_s^{-1}$.  

We assume that 
%an atomic ensemble of atoms interacting with SP short pulses and  
all the atoms initially are prepared in their ground states
and move to the classical description of the light field (replacing $\hat{a}(x,t)$ by $a(x,t)=\langle\hat{a}(x,t)\rangle$).
Solving in this case the Bloch equations \eqref{Max_Bloch_eqs} 
%for $S_{z}(\Delta=0, x,z,t)=- \cos \left(|\mathcal{R}(z)|\eta(x,t)\right)$ 
we obtain 
%for the field envelope area $\eta(x,t)$ 
the following equation for the envelope area  $\eta(x)=\int_{-\infty}^{\infty}a(x,t)dt$, where $a(x,t)=\langle\hat{a}(x,t)\rangle$
(see also \cite{Allen75, Moiseev_PRA2024,Moiseev23}): 

\fla{
\left(\frac{\partial}{\partial x} + 
\frac{\gamma_w}{2v_g}\right)
\eta(x)=-\pi
\sqrt{\frac{\pi}{2}}\frac{G(0) \rho_0 }{v_g}
\times
\nonumber
\\
\int  dy dz |\mathcal{R}(z)|
\sin \left[|\mathcal{R}(z)|\eta(x)\right],
\label{integral}
}
\noindent
where 
$(\frac{\gamma_w}{v_g})^{-1}$ is a propagation length of SP pulse, we also introduced   $\eta(x,t)=\int_{-\infty}^{t}a(x,t)dt$ and $\eta(x)= \eta(x,\infty)$ and have  taken into account that 
$G(\frac{\Delta}{\Delta_{in}})=G(-\frac{\Delta}{\Delta_{in}})$, $\rho_0$ is the atomic density.
%$G(0)$ is the inhomogeneous broadening function at zero detuning, and 

The integration in Eq. \eqref{integral} is evaluated over the cross section  (y-z plane) normal to the field propagation direction $\textbf{x}$. The integral over y gives the sample lateral thickness in the y-direction, $L_y$. 

The remaining integral over z strongly depends on: (1) the two level atomic dipole orientations, (2) the mode polarization (TM or TE), (3)  the type of solutions considered (coth or tanh solutions- Eqs.6,8), (4) the location of the atomic ensemble in the graphene three layer structure and so on. 
The structure functions $F_{TM,TE}$, that we described in Eqs. \eqref{FTM-1}-\eqref{FTM+1}, determine the form of the SP-atoms coupling term $\mathcal{R}(z)$ and thus, the behaviour of solutions to \eqref{integral} and the details of its dynamics. 
We see 
%from the expressions for these structure functions 
that the coupling term $\mathcal{R}(z)$, that inters into \eqref{integral}, has the three basic forms $\text{sinh}[k(z-L/2)]$, $\text{cosh}[k(z-L/2)]$, or $e^{k(z-L/2)}$ leading to different solutions and dynamics which will be explored in the following sections. 

To keep track of the many parameters here, we shall focus discussions on two level atomic ensemble with arbitrary dipole moment orientations located between the two graphene sheets $0<z<L$ (where $L$ is the nanoscale distance) and interacting with TM polarized graphene SP modes in the asymmetric tanh solutions only. 
 
\subsection{Atomic dipole orientations}

\textbf{1. Atomic dipole parallel or normal to interface}. As seen from the TM structure functions for $0<z<L$, a dipole oriented parallel to interface (along x-direction) couples to the x-components of the TM field, while a dipole normal to interface (along z-direction) couples to the z-components of the field. So for asymmetric tanh solutions ($\zeta=-1$), the coupling term has the odd form $\text{sinh}[k(z-L/2)]$ for parallel dipole, and even form $\text{cosh}[k(z-L/2)]$ for the normal dipole. Similarly for the symmetric coth solutions ($\zeta=+1$) the couplings terms have the even $\text{cosh}[k(z-L/2)]$ and odd forms $\text{sinh}[k(z-L/2)]$.
So consider the case where the atomic dipole is coupled to the asymmetric tanh solutions modes. 
Then the coupling term is $|\mathcal{R}(z)|=\mathcal{R}_0  \text{sinh}[k_s (z-\frac{L}{2})]$ for dipole along x-direction and 
$|\mathcal{R}(z)|=\mathcal{R}_0  \text{cosh}[k_s (z-\frac{L}{2})]$ for dipole along z-direction, with $\mathcal{R}_0 = \frac{4 d_{21} M}{\hbar} \mathcal{C}_{\lambda}$. 
Now introducing the pusle area $\theta(x)=\mathcal{R}_0\eta(x)$, \eqref{integral} becomes

\fla{
\left(\frac{\partial}{\partial x} + 
\frac{\gamma_w}{2v_g}\right)
\theta(x)=- \frac{\alpha}{2}
F[\theta(x), k_s L],
\label{theta}
}
where 
the absorption coefficient $\alpha$ by atomic ensemble is

\fla{
\alpha= 8
\sqrt{2 \pi} \frac{ \omega  d_{21}^2 G(0) \rho_{0}}{\varepsilon_{0} \hbar} \left[\frac{|M|^2}{k_s L_{TM} v_{TM}}\right],
\label{Absorption-1}
}
\noindent
and we have used Eq.\eqref{Norm_factor} for $\mathcal{C}$ in the expression for the coupling term $\mathcal{R}_{0}$. 
The absorption coefficient is composed of a numerical factor that depends on the atomic ensemble and another term inside the square brackets that depends on the SP mode frequency $\omega$ and graphene separation distance $L$ as we discuss below. 

The function $F[\theta(x),k_sL]$ is given for the odd sinh mode coupling as 
\fla{
F[\theta(x), k_s L]=2 \int_{0}^{k_s L/2}   du e^{-k_s L/2}  \text{sinh}(u)
\nonumber
\\
\sin \left[\theta(x) e^{-k_s L/2} \sinh{u}\right],
\label{Fsinh-function}
}
and for the even cosh mode coupling we have a similar equation with $\text{cosh}$ function replacing the $\text{sinh}$ functions. 

\textbf{2.Atomic dipole orientation in the x-z plane}.
Again we consider atomic  dipole located within the two graphene sheets, this time with in-plane dipole moment   
$\textbf{d}_{21}=\hat{x}d_x-i\hat{z}d_{z}$.   
From the TM structure functions in \eqref{FTM-1}-\eqref{FTM+1} for $0<z<L$, this  dipole couples to both x and z components of the TM field, and by engineering the dipole moment such that $d_z=d_x (k_s/k_{||})$ we have in this case exponential coupling mode $|\mathcal{R}(z)|=\mathcal{\tilde{R}}_0 e^{-k_s z}, $  
$\mathcal{\tilde{R}}_0 = \frac{4 d_{21} M \zeta}{\hbar} \mathcal{C}$. This anzats for the dipole moment is easily implemented for example in the non-retarded (or electrostatic) limit where $k_{||}\gg\omega\sqrt{\varepsilon}/c$. This requirement is already fulfilled since the graphene SP wave number $k_{||}$ is two orders of magnitude larger than the free space wave number $\omega/c$. In this case $k_{||}\approx k_s$ and this choice leads to dipole moments $d_x=d_z=d_{21}$ which is easily met. 
For the exponential mode coupling we have the following function $F[\theta(x),k_s L)$ 

%Inserting this exponential coupling into \eqref{integral} we obtain the same Eq.\eqref{theta} for the other two sinh and cosh modes, but with absorption coefficient $\tilde{\alpha}= \alpha e^{k_sL}$, and function $F[\theta(x),k_s L)$  
\fla{
F[\theta(x), k_s L]=\int_{0}^{k_s L}   du  e^{-u}
\sin \left[\theta(x) e^{-u}\right].
\label{Fexp-function}
}

In their present forms, Eqs. \eqref{Fsinh-function} for the sinh and cosh modes coupling , and \eqref{Fexp-function} for the exponential mode coupling, admit no analytical solutions for arbitrary values of pulse area $\theta(x)=\eta(x)\mathcal{R}_0$ and graphene sheet separation $k_s L$. Approximate solutions are possible to find for weak field signals, and nanoscale distance between graphene sheets, which we discuss next. 
The small distance $L$ and weak light case are particularly interesting in the context of creation of compact quantum memory (QM) cell, since the analysis of the conditions for the implementation of a compact QM cell is based on the consideration of the interaction of extremely weak light or SP pulses with an atomic ensemble with the search for conditions for achieving its high efficiency due to the use of enhanced interaction of radiation with atoms.
%the analysis of the conditions for the implementation of effective QM is based on the consideration of the interaction of extremely weak light or SP pulses with an atomic ensemble with the search for conditions for achieving its high efficiency, which is possible for sufficiently strong interaction.

\subsection{Weak SPs fields}

We consider the evolution of weak graphene plasmonic pulses characterized by a small pulse area  ($\theta(x=0)\ll \pi$)  in the three coupling mode cases under consideration. Eq.\eqref{Fsinh-function} gives in this limit

\textbf{1.}  $\text{sinh}[k_s(z-\frac{L}{2})]$-mode, 
\fla{
F[\theta(x), k_s L]= 
 \theta(x) \left[\frac{1}{2}\text{sinh}(k_s L) -\frac{1}{2} k_s L\right] e^{-k_s L}.
\label{Fsinh-function-1}
}

\textbf{2.}  $\text{cosh}[k_s(z-\frac{L}{2})]$-mode

\fla{
F[\theta(x), k_s L]= 
 \theta(x) \left[\frac{1}{2}\text{sinh}(k_s L) + \frac{1}{2}k_s L\right] e^{-k_s L},
\label{Fcosh-function-2}
}
while \eqref{Fexp-function} gives
 
\textbf{3.} $e^{-k_s z}$-mode
\fla{
F[\theta(x), k_s L]= 
\theta(x)
  \left[\text{sinh}(k_s L)
 e^{-k_s L}\right].
\label{Fexp-function-3}
}

 \begin{figure}
\includegraphics[width=1.0\linewidth]{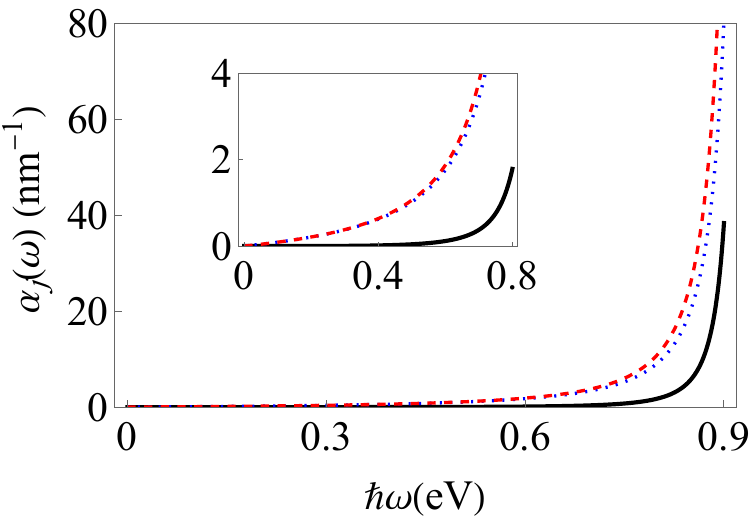}
    
\caption{ Effective absorption coefficients in the weak field limit given as functions of the mode energy $\hbar\omega$ for graphene sheet separation $L=2nm$. The red dashed line is the exponential mode absorption $\alpha_\text{x}$, the blue dotted line is the cosh mode absorption $\alpha_c$, and the black solid is the sinh mode absorption $\alpha_s$. The absorption coefficients are of the order $1 nm^{-1}$ comparable to the wave number $k_{||}=\text{Re}K_{||}$, and 2-3 orders of magnitudes larger than losses $\text{Im}K_{||}$. Parameters used are $d_{21}=5\times10^{-32} C.m,G(0)=2\times10^{-8}s, \rho_0=10^{25}m^{-3}, \varepsilon_1=1,\varepsilon_s=2, E_F=1eV, \hbar/\tau=0.001 eV$, which are typical of rare-ions at helium temperature (see \cite{GOLDNER-2015}).}
\end{figure}

\begin{figure}
\includegraphics[width=1.0\linewidth]{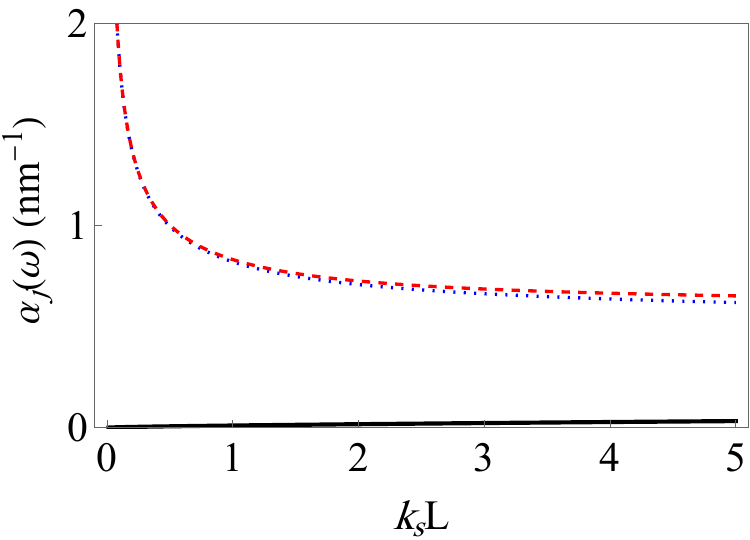}
\caption{Effective absorption coefficients in the weak field limit given as functions of the scaled sheet separation $k_s L$, for mode frequency $\hbar\omega=0.5 E_F$ and $k_s=10^{9}m^{-1}$. The red dashed (upper) line is the exponential mode absorption $\alpha_\text{x}$, the blue dotted line is the cosh mode absorption $\alpha_c$, and the black solid is the sinh mode absorption $\alpha_s$. As in Fig.4. the absorption coefficients are of the order $1$ $nm^{-1}$.  Parameters are as in Fig.4.}

\end{figure}

Inserting the approximate expressions Eqs.\eqref{Fsinh-function-1}-\eqref{Fexp-function-3} into expression \eqref{theta} we see that all three modes have same linear equation for the pulse area and follow the Beer-Lambert exponential decay law with distance \cite{Allen75} with three effective  absorption coefficients sharing the same multiplying factor $\alpha$, and an envelope function for each absorption coefficient given by 
$\alpha_s=\frac{\alpha}{2}[\sinh{(k_s L)}-(k_sL)]$ 
for the sinh mode coupling, 
$\alpha_c=\frac{\alpha}{2}[\sinh({k_s L})+(k_sL)]$ 
for the cosh mode coupling, 
and $\alpha_\text{x}=\alpha\sinh({k_s L})$ for the exponential mode coupling. 
The behaviors of the three effective absorption coefficients are shown in Fig.4 and Fig5. In Fig.4 we show the absorption coefficients as functions of the SP mode frequency $\hbar\omega (eV)$ (or the atomic transition frequency $\omega_0$, since we consider resonant interaction) for fixed graphene sheet thickness $L=5 nm$, while Fig.5 shows absoprtion coefficents as function of the scaled graphene sheet separation $k_sL$ for fixed mode (or atomic transition) frequency $\hbar\omega_0=0.5 E_F$ and $k_s=1nm^{-1}$. In these two figures, the red dashed (upper) line is the exponential mode absorption $\alpha_\text{x}$, the blue dashed line is the even absorption $\alpha_c$, and the black solid is the odd sinh mode absorption $\alpha_s$.
The magnitudes of absorption coefficients are of the order $\alpha_j=1nm^{-1}$ comparable to parallel wave number $k_{||}=\text{Re}(K_{||})$ and 2-3 orders of magnitudes larger than losses $\text{Im}(K_{||})$, which justifies the assumption $\alpha\gg\gamma_w/2v_g$.

The small sheet separation $k_s L<1$ behavior is strongly modified and modulated by the factor in the square brackets of $\alpha$ which is itself a function of $k_sL$, and in particular the group velocity $v_g$ , which is shown in the inset of Fig.5. This factor has the graphene structure finger prints and accounts for the modification of absorption coefficients due to graphene SP modes. For large separation the absorption coefficients $\alpha_s,\alpha_c$ saturate to the value $\alpha/4$ while the coefficent $\alpha_x$ saturates to the value $\alpha/2$. 

Another important issue in the development of QM for plasmons is the search for its implementation in the most compact spatial scheme, for example in the nanoscale regime. To analyze the possibilities of implementing such schemes, we consider next the graphene structure with small sheet separation.

\subsection{Small sheet separation} 

In this case, the atomic ensemble is localized between graphite sheets separated by a nanoscale distance satisfying the condition $k_s L \leq 1$. 
%In this case, the atomic ensemble is localized in the smallest volume, 

\textbf{1.} For the $\text{sinh}{[k_s (z-L/2)]}$-mode, 
at small distance between graphene sheets   $k_s L/2\leq$ , but arbitrary pulse area $\eta(x)$, 
$\text{sinh}[u]\cong u$.
In this case we have from \eqref{Fsinh-function}:

\fla{
&F(\theta_1(x),k_s L)= \frac{1}{\mathcal{R}_0^2 \eta(x)^2}\int_{0}^{\theta_1(x)} dZ \cdot
Z  \sin \left( Z\right)
\nonumber \\
=& \frac{1}{\mathcal{R}_0^2 \eta(x)^2}
\left[\sin \theta_1(x,L) - \theta_1(x,L) \cos \theta_1(x,L)\right],
\label{F-function-dop-2}
}
where $\theta_1(x,L)= \eta(x)\mathcal{R}_0 k_s  L/2$ 
is an effective pulse area, which summarizes the response of atoms located between graphene sheets.
Substituting this expression into \eqref{theta}, we have the following equation for the pulse area

\fla{
\left(\frac{\partial}{\partial x} + 
\frac{\gamma_w}{2v_g}\right)
\theta_1(x,L)=- \frac{\alpha_1}{2}
F_1[\theta_1(x,L)],
\label{theta-1}
}

%where $\alpha_1= \pi
%\sqrt{\frac{\pi}{2}}\frac{G(0)}{v_g} \rho_0 \mathcal{R}_0 k_s L^2\cdot \mathcal{R}_0 k_s L/2$
\noindent
where the function

\fla{
F_1[y]=3\frac{\sin y - y \cos y}{y^2},
\label{F-theta1}
}
and the absorption coefficient $\alpha_1$  takes the form 
\fla{
\alpha_1=\pi
\sqrt{\frac{\pi}{2}}\frac{G(0)}{3! \cdot v_g} \rho_0 (L_y L) (\mathcal{R}_0 k_s L)^2,
\label{absorb_coef_1}
}

\noindent
showing its growth $\sim L^3$.
The function $F_1[\theta_1(x,L)]$ is depicted in Fig.6, which is characterized by decaying oscillations with increasing  $\theta_1(x,L)$.
In the limit of weak input radiation described by a small pulse area $\theta_1(x,L)\ll1 $, we obtain $F_1[\theta_1(x)]\cong \theta_1(x,L)$.
Such weak radiation will be absorbed by atoms at a distance of an order $(\alpha_1+\frac{\gamma_w}{v_g} )^{-1}$ according to the Eq. \eqref{theta-1}, forming a dephasing  collective coherence in atomic ensemble \cite{Moiseev_FF2025}.

%\fla{\alpha_1=&\frac{\alpha}{3}\left(\frac{k_s L}{2}\right)^3=\nonumber \\
%&\pi \sqrt{\frac{\pi}{2}}\frac{G(0)}{3! \cdot v_g} \rho_0 (L_y L) (\mathcal{R}_0 k_s L)^2,
%\label{absorb_coef_1}}

%and the function $F_1[\theta_1(x,L)]$ on the right side of the Eq.\eqref{theta-1-xL} takes the form:
%$\alpha_1=\frac{\alpha}{3}\left(\frac{k_s L}{2}\right)^3=$
%$\pi \sqrt{\frac{\pi}{2}}\frac{G(0)}{3! \cdot v_g} \rho_0 (L_y L) (\mathcal{R}_0 k_s L)^2$

%$\alpha_1= \pi \sqrt{\frac{\pi}{2}}\frac{G(0)}{3 v_g} \rho_0 (L_y L) (\mathcal{R}_0 k_s L)^2$ 
%is an absorption coefficient in this case,

%\fla{F_1[\theta_1(x,L)]=3\frac{\sin \theta_1(x,L) - \theta_1(x,L) \cos \theta_1(x,L)}{\theta_1^2(x,L)}.
%\label{F-theta-1-xL}}

As the effective pulse area increases, the function $F_1[\theta_1(x,L)]$ exhibits nonlinear oscillatory behavior, which strongly affects the solution of the Eq.\eqref{theta-1}.
There are two nearest critical values of the pulse area
%There are two critical values of the 
$\theta_1(x=0,L)=(\Theta_1,\Theta_2)$ ($\Theta_1\cong 1.43 \pi$, $\Theta_2\cong 2.46 \pi$) in   the $F_1[\theta_1(x=0,L)]$.
 The first root $\Theta_1 $  corresponds to the first critical point $\pi$ of unstable equilibrium  in standard area theorem \cite{MacCall69} and $\Theta_2$  corresponds to the point  $2\pi$ of stable evolution of pulse area, which is similar to the formation of an optical soliton in free space scheme \cite{Allen75}.
 We are interested in the case of a resonant transition with high optical density  $\alpha_1\gg \frac{\gamma_w}{v_g}$, where it is possible to neglect effects of    
%Considering the evolution of the pulse area $\theta_1(x)$ in Eq. \eqref{eta-alpha_1} for
weak losses ($\frac{\gamma_w x}{2v_g}\ll1$). 
In this case, if the input pulse area $\theta_1(x=0,L)>\Theta_1$ it evolves to the stable state $\theta_1(x,L)_{x\gg \alpha^{-1}_1}\rightarrow \Theta_2$.

 \begin{figure}
%\begin{center}\vspace{1cm}
\includegraphics[width=1.0\linewidth]{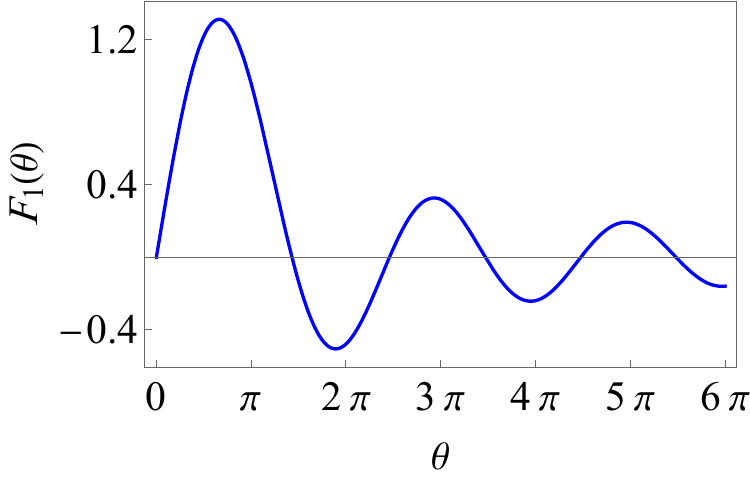}
\caption{The function $F_1(\theta)$ given by Eq.(47). There are two nearest critical values of the pulse area
%There are two critical values of the 
$\theta_1(x=0,L)=(\Theta_1,\Theta_2)$ where $\Theta_1\cong 1.43 \pi$, and $\Theta_2\cong 2.46 \pi$, in the $F_1[\theta_1(x=0,L)]$ at initial pulse area $\theta_1(x=0)$ and arbitrary sheet separation L.
 }.
\label{F-theta}
\end{figure}

It is worth noting that $\Theta_{1,2}$ are fractional values of $\pi$ and $\Theta_2$ is larger than $2\pi$,  which is due to the presence of an inhomogeneous electric field in the cross section of the graphene waveguide.
%Why the steady pulse area is greater than pi is an interesting question, the clarification of the physical cause of which requires additional analysis. Since the deviation from 2 pi is caused by the inhomogeneity of the electric field in the cross-section of the waveguide, 
The emergence of a stable pulse area solution can be explained by the fact that $\Theta_2$ corresponds to dominant part of the atoms located in the region between the graphene sheets where the electric fields are smaller than on the graphene surface, and for them the pulse area actually turns out to be $2\pi$.

It can be expected that a precise consideration of the pulse area evolution (deviation from the used approximation $\text{sinh}[y]\cong y$) will lead to a refinement of $\Theta_2$.
At the same time, taking into account that for a very large value of $k_s L\gg1$, where the functions $|\mathcal{R}(z)|$ are converted to $e^{-k_sz}$,
%At the same time, it must be borne in mind that with a very large value $k_s L\gg1$, where the functions $|\mathcal{R}(z)|$ transform into $e^{-k_s x}$, 
and the equation for the pulse area of surface plasmons does not provide stable solutions \cite{Moiseev_PRA2024}.
Finding out the reasons for the disappearance of the stable point requires further clarification of the physical nature of this behavior.
%It is important to keep in mind that a stable value of the pulse area over long distances ($\theta_1(x,L)=\Theta_2$) cannot lead to the appearance of graphene plasmon pulses with a stable temporal shape, since not all atoms will experience excitation by $2\pi$ pulses. 
It is important to keep in mind that a stable  value of the pulse area at large distances ($\theta_1(x,L)=\Theta_2$) may not lead to the appearance of graphene plasmon pulses with a stable temporal shape if some atoms do not return to the ground state after the end of the interaction.

Therefore, it can be expected that the propagation of the graphene plasmon pulse will be accompanied by an increase in its duration,  which is necessary to preserve its pulse area $\Theta_2$,  
%and a constant outflow of energy into the atomic system.
%In this case, the temporal duration of this pulse will continuously increase, which is necessary to preserve its pulse area $\Theta_2$,
but this will lead to increased relaxation and eventually to its decay.

%However, finding out the possibility of forming a stable temporal shape of fractional graphene plasmon modes is possible and  requires additional study.

\textbf{2).}
For the $\text{cosh}[k(z-L/2)]$ mode, we have $\text{cosh}[k_s (z-\frac{L}{2})]\cong 1$, that leads to constant coupling $|\mathcal{R}(z)|=\mathcal{R}_0$ which results in the usual pulse area theorem \cite{MacCall69}
\fla{
\left(\frac{\partial}{\partial x} + 
\frac{\gamma_w}{2v_g}\right)
\theta_2(x)=- \frac{\alpha_2}{2}
F_2(\theta_2(x)) =- \frac{\alpha_2}{2}
\sin \theta_2(x),
\label{eta-alpha_1-2}}
with well-known solution $\theta_2(x)=2\arctan\{\tan\left({\frac{\theta_1(0)}{2}}\right)e^{-\alpha_2 x/2}\}$
(if $\frac{\gamma_w x}{2v_g}\ll 1$), 
where $\theta_2(x)= \eta(x)\mathcal{R}_0$, $\alpha_2 =2\pi \sqrt{\frac{\pi}{2}}\frac{G(0)}{v_g} \rho_0 L_y L \mathcal{R}_0^2$.
This solution 
%of Eq.\eqref{eta-alpha_1-2} 
indicates the appearance of stable $2\pi$ plasmon graphene pulses, which reproduces the McCall-Hahn solution  
%Such propagation of the graphene plasmonic mode, based on providing a 
for uniform electric field throughout the entire atomic ensemble, which requires the creation of bilayer graphene with the necessary high control of the very small distance between its sheets.
In comparison with the sinh mode, the absorption coefficient $\alpha_2$ becomes very large $\frac{\alpha_2}{\alpha_1}=\frac{12}{(k_s L)^2}>48$ and  higher initial pulse area $\frac{\theta_2(x=0)}{\theta_1(x=0)}=\frac{2}{k_s L}>4$ for $k_sL=1/2$ that provide stronger nonlinearity in  the propagation of symmetric graphene surface mode.

We close this subsection by commenting on the practical realization of $k_s L\leq 1$. 
Taking into account the  GSP wavenumber shown in Fig.3 , in the frequency range $\hbar\omega\approx 0.5 $ eV,  $k_s\approx 5\cdot 10^8 m^{-1}$ is of interest.
Here $k_s L \approx 1 $ for distances $L \approx 2$ nm.
%This means that at least one layer of two-level atoms can be positioned  between two sheets  of graphene.
However, the use of atoms located in the closest proximity to the graphene sheet is not desirable due to their too strong interaction with the free electrons of graphene,  which causes relaxation of atomic coherence \cite{Gomez2019}.

\section{Fractional pulses between graphene sheets modes}
The exponential modes admit interesting solutions that we discuss here separately. In this case, the amplitude of the electric SP mode decreases exponentially with distance from the graphene sheets. The pulse propagation equation for the this mode is given by \eqref{theta} with function F given by \eqref{Fexp-function}, which upon integration gives the equation
\fla{
\left(\frac{\partial}{\partial x} + 
\frac{\gamma_w}{2v_g}\right)\theta_3(x)=
-\alpha
F_3[\theta_3(x)],
\label{Theta-equation-LA}
}
where 
\fla{
F_3(\theta_3)=\frac{\sin^2 [\frac{\theta_3}{2}]-\sin^2 [\frac{\theta_3}{2}e^{-k_s L}]}{\theta_3},
\label{F-3}
}

%\fla{F_3(\theta(x))=\frac{\sin^2 [\frac{\theta(x)}{2}]-\sin^2 [\frac{\theta(x)}{2}e^{-k_s L_a}]}{\theta(x)}.
%\label{F-3}}

\noindent
where $\theta_3(x)=\mathcal{R}_0\eta(x)$, and atoms are located between the graphene sheets. The  single interface case where atoms are uniformly distributed in space was considered recently \cite{Moiseev_PRA2024} and obtained in the limit $e^{-k_s L}\rightarrow 0$ 
%, where the equation for the pulse area of plasmon modes was obtained:
%\fla{
%\left(\frac{\partial}{\partial x} + 
%%\frac{\gamma_w}{2v_g}\right)\theta(x)=
%-\alpha
%\frac{\sin^2 [\theta(x)/2]}{\theta(x)},
%\label{Theta-equation}
%}
leading to the analytical solution

 \fla{
x=x_0-
\frac{1}{\alpha}\left(T[\frac{\theta(x)}{2}] 
-T[\frac{\theta(x_0)}{2}]\right),
\label{Solution}
}
\noindent
where
\fla{
T[y]=\text{ln}[\sin\left(y\right)]-y\cot\left( y\right),
\label{T-function}
}

\noindent
the pulse area $\theta(x)=\theta_3(x)=\mathcal{R}_0\eta(x)$ and absorption coefficient $\alpha$ \cite{Moiseev_PRA2024}.
The analytical solution \eqref{Solution} shows the formation of $2 \pi$ pulses, which, however, are unstable and begin to weaken intensively with a slight decrease in the pulse area relative to $2 \pi$. 

Although it is difficult to get an exact analytical solution to Eq. \eqref{Theta-equation-LA} for arbitrary distance $L$ and pulse area,
nevertheless it is possible to find the formation of steady-state propagation of surface plasmon modes for a number of interesting cases.

 Let us assume $e^{-k_sL}=\frac{1}{2}$. Here we get

\fla{
F_3[\theta(x)]=2 \frac{\sin^2 [\frac{\theta(x)}{4}](\cos [\frac{\theta(x)}{2}]+\frac{1}{2})}{\theta(x)}.
}

Substituting this function into the Eq. \eqref{Theta-equation-LA}, we find that its solution indicates the presence of stable points of evolution for the following  fractional values $\Theta_2=2\pi+\frac{2}{3}\pi+4n\pi$ (n=1,2,...). 
The same solutions are obtained for $e^{-k_sL}=\frac{1}{4}$.
Unstable evolution points,  similar to solutions of Eq. \eqref{Solution}  exist for $\Theta_1=2m\pi$ (m=1,2,...).  
%This function gives unstable evolution points, similar to solutions of Eq. \eqref{Solution}  but with different values of  $\Theta_1=2m\pi$ (m=1,2,...) and stable solution similar to the case 1, and also with different fractional values $\Theta_2=2\pi+\frac{2}{3}\pi+4n\pi$ (n=1,2,...). 
%$ii)$. 
This solution can be generalized directly to the  cases $e^{-k_sL}=\frac{1}{2^m}$, 
where we find stable solution with fractional values $\Theta_2=\frac{2^{m+1}}{3}\pi$.
%Such  of the pulse area are interesting general feature of these solutions. 
Thus, a decrease of $e^{-k_sL}$ requires an increase of the pulse area $\Theta_2$.
This is explained by the need to achieve a pulse area of $2\pi$ for atoms having a weaker coupling constant with the graphene mode.
We can conclude that the presence of a fractional value $\Theta_2$ (not exactly equal to $2n\pi$ where $n=1,2,...$) reflects the presence of continuous distribution
%We can conclude that since $\Theta_2$ does not equal $2n \pi$ ($n=1,2,...$) reflects again the presence of a continuous distribution 
of the coupling constant $|\mathcal{R}(z)|$ across the atoms of the medium.

\begin{figure}%
    \centering
    \subfloat {(a){\includegraphics[width=8cm]{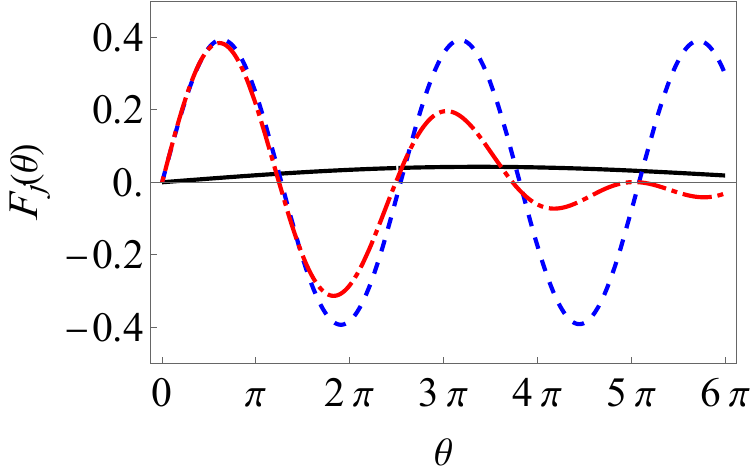} }}%
    \qquad
    \subfloat{(b){\includegraphics[width=8cm]{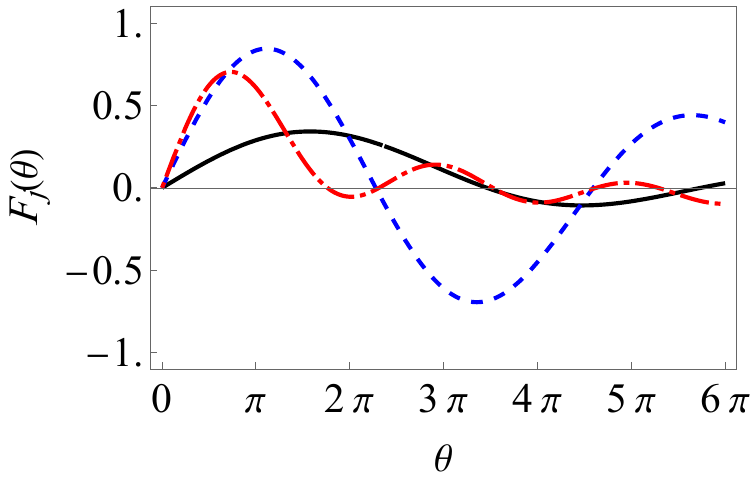} }}%
    \caption{
    The exact functions $F(\theta,k_sL)$ for the three coupling modes as in Eqs.\eqref{Fsinh-function}-\eqref{Fexp-function} as functions of the pulse area $\theta(x)$. The black solid line  corresponds to sinh coupling mode, blue dashed line - to cosh coupling mode and red dashed-dotted line - to the exponential coupling mode. Graphene sheet separation $k_{s}L=0.5$ (a) and $k_{s}L=2$ (b). Critical points occur at $\theta_n\approx 2\pi+\pi/2$ for blue-dashed and red (dashed-doted) (a), and $\theta_n\approx 2\pi+3\pi/4$ for (b).}%
    \label{fig:example}%
\end{figure}

\begin{figure}%
    \centering
    \subfloat {(a){\includegraphics[width=8cm]{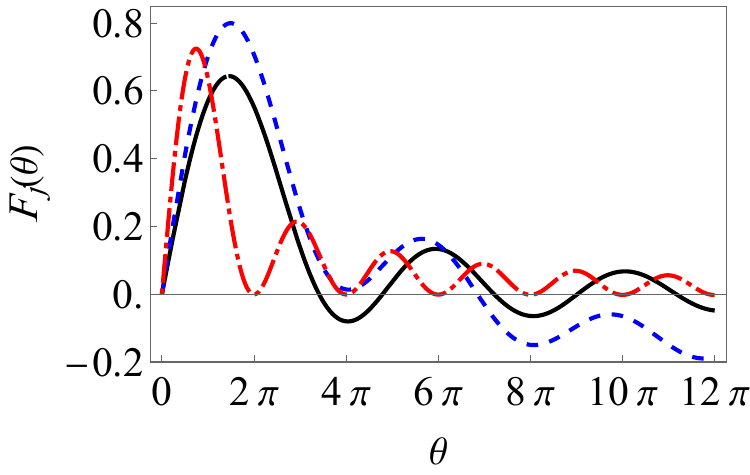} }}%
    \qquad
    \subfloat{(b){\includegraphics[width=8cm]{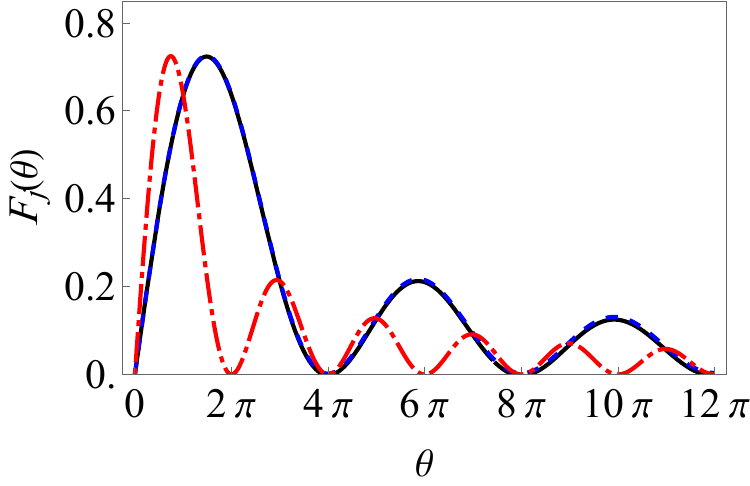} }}%
    \caption{The exact functions $F(\theta,k_sL)$ as in Fig.7, but for graphene sheet separation $k_{s}L=4.5$ (a) and $k_{s}L=9$ (b). No stable solutions for this latter case}%
    \label{fig:example}%
\end{figure}
\begin{figure}
%\begin{center}\vspace{1cm}
\includegraphics[width=1.0\linewidth]{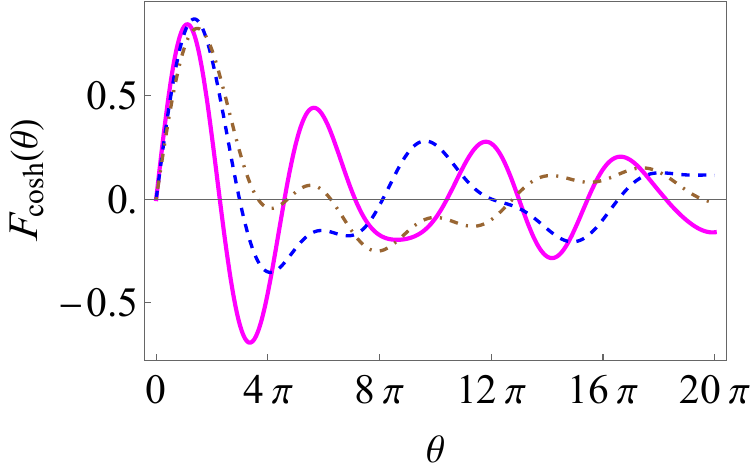}
\caption{The exact function $F_{cosh}(\theta,k_sL)$ as in Eq.\eqref{Fsinh-function} but for cosh coupling mode with sheet separation $k_{s}L=2$ -the magenta solid line;  $k_sL=3$ - blue dashed line;  $k_sL=4$ - brown (dashed-dotted) line;
$\theta_n\approx 15\pi+\pi/2$ for magenta solid  line, 
$\theta_n\approx 16\pi+\pi/2$ for blue solid curve, $\theta_n\approx 13\pi$ for brown (dashed-doted) curves.}
\label{Fig-6}
\end{figure}

The solutions obtained above, showing the fractional value of the stable pulse area $\Theta$ of plasmons, also occur with other interaction parameters.
In Figs.7-9, we show  the behaviour of the exact functions $F$, corresponding to the cases of sinh, cosh and exponential coupling modes as given in \eqref{Fsinh-function}-\eqref{Fexp-function}.  
For comparison, we present four series of graphs calculated for  different distances: $k_s L=0.5, 2, 5$ and $9$.  
From these graphs, it can be seen that increasing the distance leads to a weakening of the oscillations. 
It is seen that if the distance is sufficiently small, for example, for $k_s L=0.5$ and  $k_s L=2$, the graphs have points $\theta=\theta_n$ at which the functions  $F(\theta_n,k_sL)=0$  and have a positive derivative $\partial_{\theta} F(\theta,k_sL)_{\mid\theta=\theta_n}>0$, which ensures the existence of a stable evolution of SP mode in Eq.\eqref{integral}.
As it can be seen from the graphs in Figs. 7 and 8, all these points  $\theta_n$ of stable evolution of SP pulses have a fractional value, which depends on the magnitude of $k_sL$.
At the same time, as the distance $k_sL$ increases, all the oscillations shift to a positive upper half-plane. The exponential coupling mode graph is shifted first, which is noticeable already at $k_sL=4.5$ (see Fig.9).
At $k_sL=9$ this is also evident for sinh and cosh coupling modes (see Fig.9), 
where only unstable solutions are possible in the propagation of SP-pulses.
It can be seen from Fig. 9, that the numerical solution in this case for the sinh and cosh coupling modes differs from the solution for the exponential coupling mode having an analytical solution only by a scale factor of 2. 
Such behavior of these SP modes is also observed for large values of pulse area $\theta$, which indicates the presence of a similar analytical solution for sinh and cosh coupling modes in the limit $k_sL \gg 1$, the search for which by analytical methods is an independent task.

Comparison of the obtained solutions and graphs in Figs. 7-9, we can conclude that the cosh coupling mode can provide the existence of stable SP pulse  distances from small distance $k_sL=0.5$ to $k_sL=4$ (see Fig.9, $\theta_n \approx 13\pi$ for the case with $k_sL=4$) with considerable enhancement of the interaction with an atomic ensemble.
However, for longer distances ($k_sL\geq 9$), our numerical simulations show that there are no stable propagation of the SP mode.
The  modes of interaction of GSP pulse with resonant atomic ensembles discussed above, which demonstrate strong absorption of signal  GSP pulses by atomic ensembles, are interesting to analyze in more detail in the context of their use in the implementation of compact quantum memory.
%\textcolor{red}{discuss. This concludes the general analysis of SP mode propagation in a two-layer graphene containing a resonant atomic ensemble}.
%In Fig.10, obtained at $k_sL=5$, we also see that the exponential mode (indicated in red) provides only unstable solutions.

\section{Quantum memory for graphene surface modes}

We have alluded in previous discussions to the concept of nanoscale quantum memory (QM) and its implementation using interaction of graphene surface modes with inhomogeneously broadened atomic ensembles.
This type of QM can use various protocols \cite{MoiseevPRL2025}of photon (spin) echo effect, which ensures the reversible dynamics in behavior of both atomic (spin) ensembles and resonant fields interacting with them \cite{Moiseev_FF2025}. 
In this section we briefly outline how to demonstrate the possibility of reversible dynamics in the proposed scheme, assuming negligible relaxation and Langevin forces during the interaction of atoms with SP pulses in Eqs. \eqref{Max_Bloch_eqs}.

%To demonstrate the reversibility of the dynamics of the interaction of atoms and plasmon modes, we focus on the  negligible weak relaxation and Langevin  forces in Eqs.\eqref{Max_Bloch_eqs}.
As we analyzed above (see discussion after Eq. \eqref{F-theta}), using these equations, weak graphene plasmon field are absorbed by the atomic ensemble when a collective quantum coherence is excited in the atoms, which has the form of a traveling wave with a wave vector $\textbf{k}_{||}$.
The inhomogeneous broadening of the atomic transition causes disappearance of the macroscopic atomic polarization, which leads to the end of their coherent interaction with the SP pulse and the plasmon field goes into a vacuum state.
In particular, we note that in controlling the reversible dynamics of the QM circuit under consideration, additional pulses resonant to adjacent atomic transitions can be used, as is the case in the CRIB fas protocol. 
In this case, when using a three-level atomic optical quantum coherence system, it becomes possible to transfer to the lower pair of spin sublevels, which significantly increases the lifetime of the quantum memory.

The possibilities of restoring the input SP pulse can be demonstrated by using a CRIB-protocol \cite{Moiseev_FF2025} in which the frequency offsets in atomic ensembles is inverted  ($\Delta_j\rightarrow - \Delta_j$) and the SP wave vector in the atomic coherence atom is reversed $\textbf{k}_{||} \rightarrow - \textbf{k}_{||}$.
Changing the sign of the frequency offsets  will lead to the restoration of the collective atomic coherence that disappeared due to dephasing during its excitation by the input signal SP pulse. 
The equations of motion describing the emission of an echo signal $\tilde{a}_e(x,t)$ radiated in the backward direction to the input signal will have the form \cite{MoiseevPRL2025}

\fla{
&\left( \frac{\partial}{\partial t}- v_g \frac{\partial}{\partial x} \right) \hat{a}_e(x,t) =
\nonumber
\\
&i\sqrt{\pi/2} \sum_{j} \mathcal{R}^{*}(z_j) \hat{\sigma}_{-}^j(t)\delta(x-x_j),
\label{Max_Bloch_eqs-echo}
\\
&\frac{\partial \hat{\sigma}_{-}^j(t)}{\partial t} =   i\Delta_{j}\hat{\sigma}_{-}^j(t)-\frac{i}{2}\mathcal{R}(z_j)\sigma_{z}^j(t)\hat{a}_e(x_j,t),
\nonumber
\\
&\frac{\partial \hat{\sigma}_{z}^j(t)}{\partial t} = i\left[
\mathcal{R}(z_j)\hat{\sigma}_{+}^j(t)\hat{a}_e(x_j,t)
-\mathcal{R}^*(z_j)\hat{\sigma}_{-}^j(t)\hat{a}^{\dagger}_e(x_j,t)\right].
\nonumber
}

Comparing this system of Eqs. \eqref{Max_Bloch_eqs-echo} describing  the emission of the SP echo signal with  the Eqs. \eqref{Max_Bloch_eqs}, describing the absorption of the input SP signal, we note that they  merge into each other with under the transformation (with conditions of neglecting the relaxation terms in Eqs.\eqref{Max_Bloch_eqs-echo}): 
%$\Delta_j\rightarrow -\Delta_j$, 
$t\rightarrow-t$, 
$\hat{a}_e(x,t)\rightarrow -\hat{a}(x,t)$ and $\hat{a}^{\dagger}_e(x,t)\rightarrow -\hat{a}^{\dagger}(x,t)$.
This means that the emission of the echo signal will be reproduced in the reverse order of time with respect to the input signal absorption process.
It is noteworthy that the echo signal receives an additional $\pi$ phase, which reflects the transition from the signal absorption process to the induced process of echo pulse emission.

Effective retrieval of the signal SP pulse will take place in the case of complete absorption of the input SP pulse by the atomic ensemble ($\alpha x_0\gg 1$, in particular  $\alpha x_0\approx 5$), where $x_0$ is a  linear dimension along the $x$-axis, inside which resonant atoms are located. 
Similar analysis \cite{Kraus2006,Moiseev2013,MoiseevPRL2025} of the optical echo protocols of QM were carried out earlier, where more general conditions for the realization of reversibility of the dynamics of the interaction of radiation and a system of resonant atoms were found.
For a weak input SP pulse, the strongest absorption will occur for exponential and cosh coupling modes in the graphene structure with a sufficiently small distance between the sheets $k_s L <10 $ (see Fig.5).
Taking into account the numerical calculations of absorption coefficients shown in the Fig.5, we estimate $ x_0\approx 5- 25$ nm, which indicates a rather small size of QM cell.

We further note that the CRIB protocol and its implementation through the use of external electric or magnetic field gradients (GEM-protocol \cite{hetet2008electro,moiseev2008efficiency}) provide an ideal scenario for reversible dynamics.
In practice, its implementation becomes complicated if it is necessary to work with broadband signal fields.
In this case, other variants of inhomogeneous  broadening are of interest, for which their own methods for controlling the phasing of atomic coherence are being developed in ROSE-, NLPE-, Hybrid-, ARM-protocols using natural inhomogeneous  broadening, AFC-protocol exploiting periodic structure of resonant lines    \cite{Chaneliere2018,Moiseev_FF2025}
and the protocol based on using macroscopic spin coherence \cite{MoiseevPRL2025}.
Despite the deviation from the exact temporal reversibility, these protocols retain the main properties of the CRIB/GEM protocol and are able to ensure high efficiency when choosing appropriate interaction parameters.
Therefore, it will be of particular interest to choose the best implementation options of various QM echo protocols for the most efficient implementation of QM  in the considered scheme of interaction of SP pulses with atoms.
%In particular, we note that in controlling the reversible dynamics of the QM under consideration, additional SP pulses resonant to adjacent atomic transitions can be used, as is the case in the CRIB-, GEM-, ROSE-, AFC protocols \cite{Chaneliere2018,Moiseev_FF2025} and in the protocol based on using macroscopic spin coherence \cite{MoiseevPRL2025}. 
The use of a three-level atomic systems is of undoubted interest where it becomes possible to transfer optical quantum coherence to the lower pair of spin sublevels, which significantly increases the lifetime of QM protocols.
%These protocols use various variants of inhomogeneous broadening of the resonance transition with adapted methods for controlling this coherence.

%At the same time, it is important to note that all these protocols are already losing strict reversibility of time dynamics, a deviation from which can lead to both a decrease in the efficiency and fidelity of signal field reconstruction. Compensation for the loss of strict temporary reversibility is one of the issues of developing effective ways to implement them \cite{Moiseev_FF2025}.

The implementation of  QM on surface plasmons considered here is notable for the fact that a QM cell can be very compact (of nanoscale dimensions) and directly integrated into a nanoscale  circuits. 
Namely, as shown above, the spatial size of a QM cell can be significantly smaller than the distances over which plasmonic qubits can propagates without noticeable attenuation. 
It is important that in this case, the lifetime of the quantum coherence in the QM scheme under consideration will be determined by the long lifetime of the quantum coherence of the atomic ensemble, rather than the surface plasmons.
At the same time, due to the strong enhancement of the interaction between the surface plasmon and a single atom, such atoms can be located near a QM cell and act as quantum processors.
The use of enhanced interactions of surface plasmons with individual atoms on the graphene surface has already become a platform for solving a wide range of problems \cite{Koppens2011}.
Such quantum processors can be various natural and artificial  atoms with a sufficiently long quantum coherence time, for example, quantum dots \cite{Zwerver_2022}, color centers in diamond \cite{pezzagna2021quantum}, superconducting qubits \cite{Huang_2020}, rare-earth ions \cite{kinos2021rare_earth_qc} etc. 
In the proposed scheme, a number of quantum devices, including quantum processors, routers, switches, waveguides and others, can be placed at a short distance from QM cell. 
The described architecture of quantum devices will be capable to perform complex quantum processing on a single graphene platform.

%A detailed study of QM operation  with quantum processors could be  the topic of special research.

\section{Conclusion}

In this work we proposed a  graphene plasmonic platform for nanoscale  manipulation and control of the interaction of quantized surface plasmons with resonant atomic ensemble.
The SP graphene modes have been determined and coupled to coherent two-level inhomogeneously broadened atomic ensemble positioned between two 2D graphene sheets separated by nanoscale  distance L. 
The relevant equations of motion have been solved analytically and numerically and physical quantities are investigated. 
In particular, we have derived an "area theorem" for the SP pulses propagating between the two graphene sheets. 
The analysis has been carried out for different dipole orientations and varying graphene sheet separations with rich set of parameters that can be used to manipulate and control the interaction of SP pulses with atomic system, namely dipole moment orientations, SP mode polarizations, type of (even or odd) solutions, locations within the graphene structure and so on. Although the analysis has been performed for atomic ensemble located within the graphene sheets, parallel analysis can be conducted for the atomic ensemble in the upper or lower outer layers.

We have identified the physical conditions necessary to create stable SP pulses propagating over long distances and predict the propagation of stable SP pulses with fractional values of pulse areas at arbitrary graphene sheet separations. 
We briefly demonstrated how the proposed scheme of double layer graphene platform  can be used to implement compact QM cell at the nanoscale level.   
We also showed that a QM cell can be co-located with several nanoscale quantum devices on the considered graphene platform, which could be interesting for implementing quantum processing with a large number of qubits.
A detailed study of how QM works with quantum processors and their control of external classical fields is the subject of ongoing field of research and will be reported in due course.

\section{ACKNOWLEDGMENTS}

 Moiseev S.A. was supported by the Ministry of Education and Science of the Russian Federation (Reg. number NIOKRT 
 125012300688-6).

\bibliography{main}

\end{document}